\documentclass[fleqn,10pt]{wlscirep}

\usepackage{times}
\usepackage{graphicx}
\usepackage{amsfonts}
\usepackage{amsmath, amsthm, amssymb}
\usepackage{dsfont}
\usepackage{mathptm}
\usepackage{color}
\usepackage{subcaption}
\usepackage{tikz}

\usepackage{sidecap}

\newcommand{\bs}[1]{{\textbf{\textit{#1}}}}

\newcommand{\rh}{\hat{\rho}_3} % Vector k
\newcommand{\gr}{\hat{g}^R} % Vector k

\newcommand{\vecS}{\bs{S}} 
\newcommand{\vecf}{\bs{f}} 
\newcommand{\tf}{{\tilde{\bs{f}}}}
\newcommand{\hx}{\hat{\bs{x}}}
\newcommand{\hy}{\hat{\bs{y}}}
\newcommand{\hz}{\hat{\bs{z}}}

\newcommand{\ie}{\textit{i.e. }}%adv : that is to say; in other words
\newcommand{\eg}{\textit{e.g. }}%[syn: f.eks., for example, for instance]
\newcommand{\etal}{\emph{et al.}}
\def\i{\mathrm{i}}

\usepackage{soul}

\title{Controlling superconducting spin flow with spin-flip immunity using a single homogeneous ferromagnet}

\author[1]{Sol H. Jacobsen}
\author[1]{Iryna Kulagina}
\author[1]{Jacob Linder*}
\affil[1]{Department of Physics, Norwegian University of
Science and Technology, N-7491 Trondheim, Norway\\
*Correspondence to jacob.linder@ntnu.no}

\begin{abstract}
Spin transport via electrons is typically plagued by Joule heating and short decay lengths due to spin-flip scattering. It is known that dissipationless spin currents can arise when using conventional superconducting contacts, yet this has only been experimentally demonstrated when using intricate magnetically inhomogeneous multilayers, or in extreme cases such as half-metals with interfacial magnetic disorder. Moreover, it is unknown how such spin supercurrents decay in the presence of spin-flip scattering.  Here, we present a method for generating a spin supercurrent by using only a single homogeneous magnetic element. Remarkably, the spin supercurrent generated in this way does not decay spatially, in stark contrast to normal spin currents that remain polarized only up to the spin relaxation length. We also expose the existence of a superconductivity-mediated torque even without magnetic inhomogeneities, showing that the different components of the spin supercurrent polarization respond fundamentally differently to a change in the superconducting phase difference. This establishes a mechanism for tuning dissipationless spin and charge flow separately, and confirms the advantage that superconductors can offer in spintronics.
\end{abstract}
\begin{document}

\flushbottom
\maketitle

\thispagestyle{empty}

\section*{Introduction}

Current research in spintronics is attracting much attention, in large part due to the pivotal role that the quantum spin degree of freedom plays in an increasingly wide class of physical systems, ranging from ultracold atoms at the micro-Kelvin temperature scale to topological insulators at room-temperature. Spin transport in superconductors \cite{MTF1970, meservey_pr,meservey_pr2, meservey_physrep}, which historically predated spin transport experiments in non-superconducting materials \cite{johnson_prl_85}, has recently re-emerged as a potential avenue for enhancing and discovering new phenomena in spintronics. Recent results are encouraging, with experiments demonstrating not only infinite magnetoresistance \cite{li_prl_13}, but also strongly enhanced quasiparticle spin lifetimes \cite{yang_nmat_10}, spin relaxation lengths \cite{quay_nphys_13}, spin Hall effects \cite{wakamura_nmat_15}, and thermoelectric currents \cite{kolenda_arxiv_15} compared with non-superconducting structures.  

Creating and manipulating spin-flow is the central feature of superconducting spintronics  \cite{linder_nphys_15, eschrig_physrep_15}. It is known that in the presence of magnetically inhomogeneous structures, such as multilayers or ferromagnets with intrinsic textures such as domain walls, spin-polarized Cooper pairs can emerge \cite{bergeret_prl_01} which thus carry not only charge but also spin supercurrents \cite{eschrig_prl_09, alidoust_prb_10, shomali_njp_11, moor, halterman}. Experimentally, it has been demonstrated \cite{keizer_nature_06, anwar_prb_10, khaire_prl_10, robinson_science_10} that such triplet Cooper pairs can carry a dissipationless charge-current through strong ferromagnets over distances far exceeding the penetration depth of conventional superconducting order into magnetic materials. This occurs precisely due to the creation of triplet Cooper pairs which are spin-polarized and thus insensitive to the pair-breaking effect of a magnetic Zeeman-field. In fact, triplet Cooper pairs were newly experimentally observed inside a conventional superconductor \cite{dibernardo_natcom_15, kalcheim_prb_15}. In very recent developments, it has been shown that intrinsic spin-orbit coupling offers an alternative avenue for generating the long-range (LR) triplet component \cite{BT13, BT14}. In that case the appearance of the LR component depends on the relationship between the spin-orbit coupling and the exchange field, with the LR triplet defined as having its spin aligned with the exchange field. This is in contrast to the short-ranged (SR) triplet component which has its spin perpendicular to the field, and is thus vulnerable to pair-breaking in the same way as conventional singlet Cooper pairs. As we will show below, these recent developments will have profound consequences for the generation of spin supercurrents in spintronics.

To date, structures with magnetic inhomogeneities such as multiple magnetic layers have been required to create long-ranged spin-supercurrents \cite{keizer_nature_06, anwar_prb_10, khaire_prl_10, robinson_science_10}. This can be experimentally challenging for several reasons, primarily because it is far from trivial to exert control over the individual layers of magnetically inhomogeneous structures, and can be complicated yet further if the magnetic layer has intrinsic texture (such as the spiral order in Ho). Here we will show that it is possible to create a spin-polarized supercurrent using just \textit{one single homogeneous magnetic element}, which eliminates the experimental complexities and heralds a new era for harnessing the dissipationless spin-flow of superconductors in spintronics. In addition, we show that this spin supercurrent does not decay even in the presence of spin-flip processes, \eg via magnetic impurities or spin-orbit impurity scattering. This spin-flip immunity is fundamentally different from spin currents in non-superconducting structures which remain polarized for the duration of the spin relaxation time. Finally, we show that the spin polarization components of the supercurrent respond qualitatively differently to a change in the superconducting phase difference $\phi$. The surprising consequence of this is that the dissipationless charge flow and spin flow can be tuned separately. In particular, both the magnitude and the polarization direction of the spin flow is controlled via the superconducting phase, offering an entirely new way to control spin transport.\\

%%%%%%%%%%%%%%%%%%%%%%%%%%%%%%%%%%%%
%% 																		FIRST MAIN RESULT																	 %%
%%%%%%%%%%%%%%%%%%%%%%%%%%%%%%%%%%%%

\section*{Spin supercurrent with a single homogeneous ferromagnet}

\begin{figure*}
\includegraphics[width=\textwidth, angle=0]{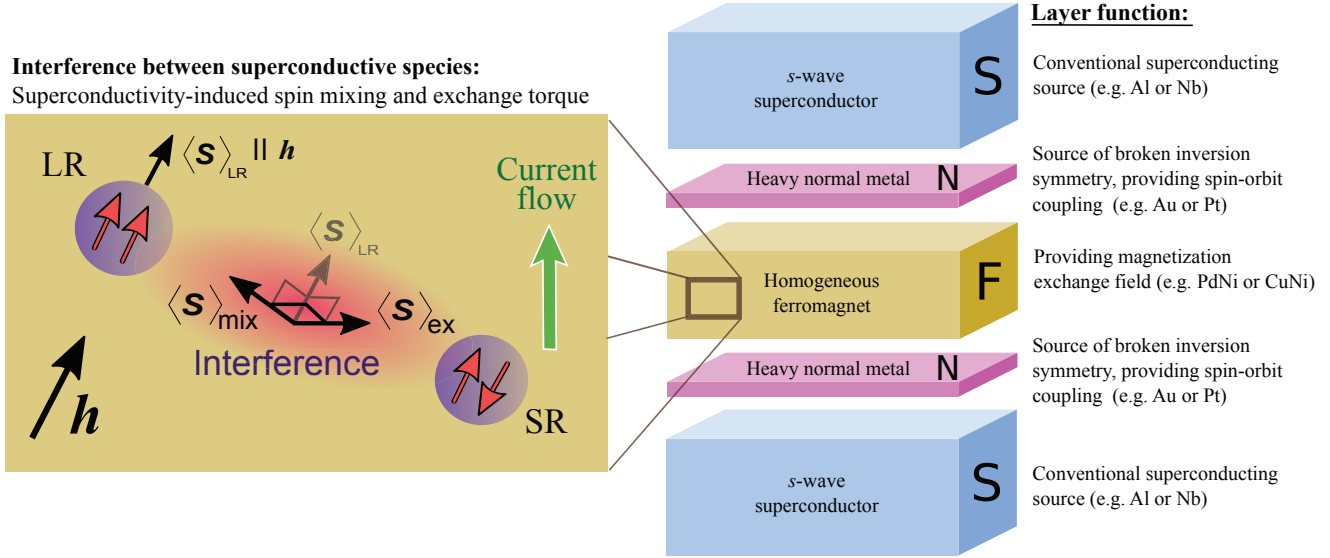}
\caption{\textbf{Proposed experimental setup and interference mechanism}. Proposed experimental setup and interference mechanism. The thin-film may be constructed with either pure Rashba spin-orbit coupling from heavy normal-metal layers (\eg Au or Pt) and a homogeneous ferromagnet with out-of-plane magnetocrystalline anisotropy (\eg PdNi or CuNi), or with both Rashba and Dresselhaus coupling in the normal layer (\eg GaAs) and a ferromagnet with purely in-plane field. In both cases, this induces an interference effect between the long-ranged and short-ranged Cooper pairs, which results in spin mixing and a novel superconductivity-mediated spin torque.}
\label{fig:model}
\end{figure*}

Consider the thin-film heterostructure depicted in Fig.~\textbf{\ref{fig:model}}, which shows a Josephson junction of conventional $s$-wave superconductive sources with normal and ferromagnetic elements typically utilized in proximity effect experiments. 

\begin{figure*}
\includegraphics[width=\linewidth, angle=0,clip]{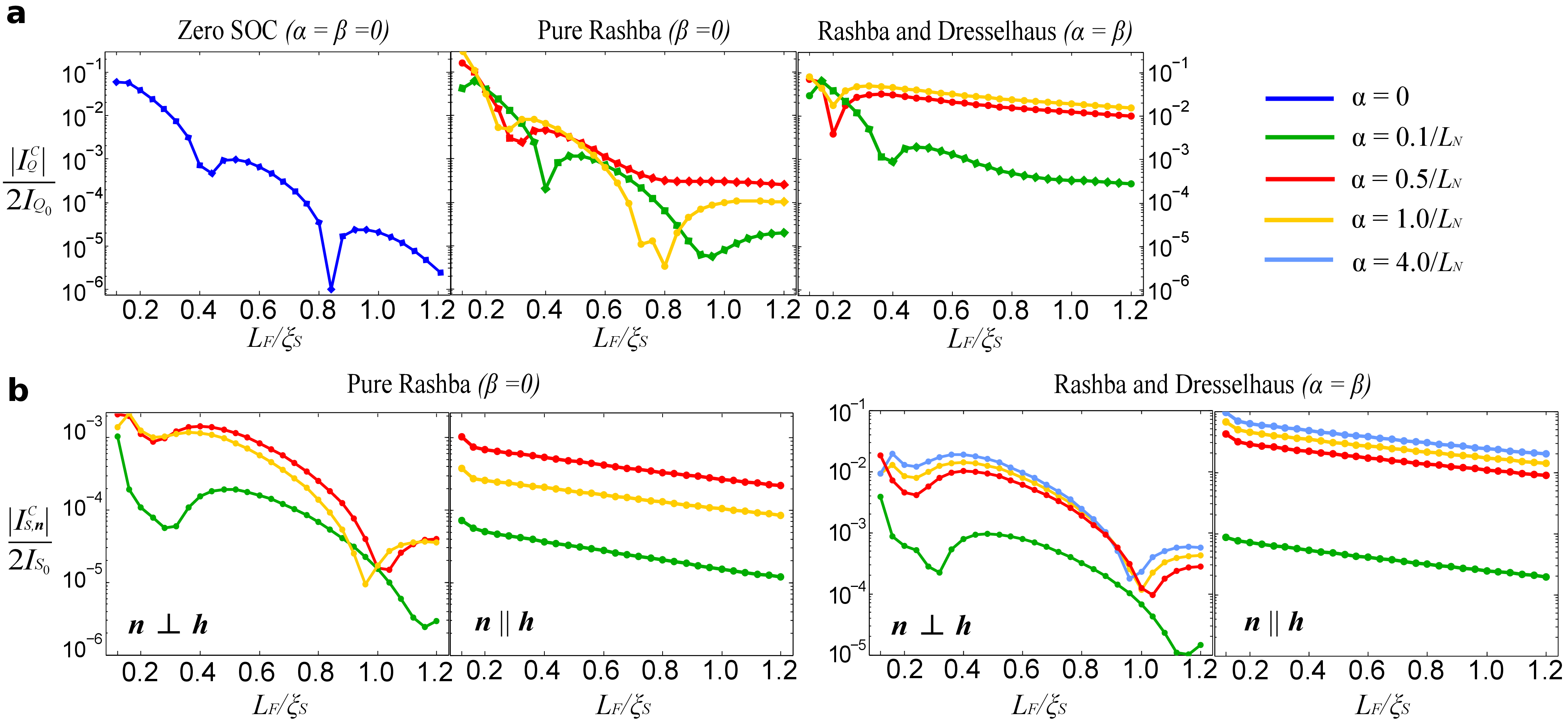}
\caption{\textbf{Charge and spin supercurrent vs. length}. The magnitude of the critical charge current $I_Q^C$ (\textbf{a}) and the components of the critical spin current $I_{S,\bs{n}}^{C}$ (\textbf{b}) in the ferromagnet as a function of the length of the layer $L_F$ is shown on a logarithmic scale. In the presence of spin-orbit coupling, the current becomes long-ranged as it makes a transition from an exponential decay with superimposed oscillations to a much slower decay with respect to $L_F$. For long ferromagnetic junctions, it is clear that the charge current is almost entirely due to the long-range component. Including both Rashba and Dresselhaus coupling results in a substantial enhancement of the critical charge currents compared with pure Rashba coupling. We assume bulk superconductivity in the superconductors, an exchange field $\bs{h}=50\Delta(0,\cos\theta,\sin\theta)$ with $\theta=0.3\pi$, and a normal metal layer length of $L_N/\xi_S=0.08$. }
\label{fig:dFCur}
\end{figure*}

We will now show that a long-ranged spin supercurrent is sustained in the junction even when only a single homogeneous ferromagnet is used. The key to achieving this is to deposit a very thin layer of a heavy normal metal such as gold or platinum at the superconducting interfaces. Recent experiments in the context of magnetization switching have shown that such interfaces will produce strong Rashba spin-orbit coupling due to the high atomic number of the metal and the interfacially broken inversion symmetry \cite{miron_nature_11}. Experimentally, care must be taken during the layer deposition in order to reduce the amount of interfacial roughness, which will in general decrease the amount of current the junction can sustain and thus affect the signal strength. The magnetic element consists of a ferromagnetic alloy which has both an in- and out-of-plane component, achievable by using \eg PdNi or CuNi, which can both feature out-of-plane magnetocrystalline anisotropy in thin-films \cite{khaire_prb_09, ruotolo_jap_04}. It is clear, therefore, that no magnetic inhomogeneities are required, and the ferromagnet does not need to feature any intrinsic spin-orbit coupling. This is in contrast to previous works that have considered long-ranged currents in either magnetically textured junctions (see \eg Refs. \cite{houzet_prb_07, eschrig_prl_09, efetov_prb_10}) or intrinsically spin-orbit coupled ferromagnets \cite{BT13, arjoranta,AlidoustHalterman2015}, where spin is not a conserved quantity, with several magnetic layers \cite{BT14}. In our setup, only a single homogeneous ferromagnet is required because the heavy normal metals supply the spin-orbit coupling, significantly reducing  the previously required level of junction complexity in order to host a spin supercurrent. Furthermore, as an alternative experimental scenario, it is possible to use a ferromagnet with a purely in-plane exchange field by employing normal layers that contain both Rashba and Dresselhaus coupling. Examples include crystals that lack an inversion structure and two-dimensional electron gases such as gallium arsenide. In this case, the singlet-triplet conversion is greatly enhanced \cite{JacobsenLinder2015,JOL}, resulting in stronger supercurrents (see Fig.~\textbf{\ref{fig:dFCur}}).

The spin-supercurrent $I_{S,\bs{n}}$ polarized along a unit vector $\bs{n}$ may be computed via the quasiclassical Green function $\check{g}$ according to\cite{alidoust_prb_10}:
\begin{align}\label{eq:spinsuper}
I_{S,\bs{n}} = I_{S_0} \int^\infty_{-\infty} \text{d}\varepsilon \text{Tr}\{ \hat{\rho}_3 \hat{\tau} (\check{g}\partial_z\check{g})^K\}.
\end{align}
Here, we have defined $\hat{\tau} = \text{diag}(\bs{n}\cdot\boldsymbol{\sigma},\bs{n}\cdot\boldsymbol{\sigma}^*)$, where $\boldsymbol{\sigma}$ is the vector of Pauli matrices, $\varepsilon$ denotes the quasiparticle energy and $K$ the Keldysh component of the Green function. $I_{S_0}=N_0\hbar D A\Delta/8L_F$, where $N_0$ is the normal-state density of states at the Fermi level, $D$ the diffusion constant and $A$ the interfacial contact area. The integral in Eq.~(\ref{eq:spinsuper}) is dimensionless since the energies have been normalized to the bulk superconducting gap $\Delta$ and lengths normalized to the ferromagnet length $L_F$. The matrix $\hat{\rho}_3=\rm{diag}(1,1,-1,-1)$. To find the Keldysh component we use the equilibrium relation
\begin{align}\label{eq:equilib}
(\check{g}\partial_z\check{g})^K = [\hat{g}^R\partial_z \hat{g}^R + (\rh \gr\partial_z\gr \rh)^\dag]\tanh(\beta\varepsilon/2),
\end{align}
where $\hat{g}^R$ denotes the retarded components of $\check{g}$ and $\beta=1/k_BT$ is the inverse temperature with $k_B$ being the Boltzmann constant. Here we have used that the advanced component of the $\check{g}$ is given by $\hat{g}^A = -(\hat{\rho}_3\hat{g}^R \hat{\rho}_3)^\dagger$. We find $\hat{g}^R$ by solving the Usadel equation for the system shown in Fig.~\textbf{\ref{fig:model}} numerically in the full proximity effect regime using the NOTUR supercomputer cluster (Kongull); see Methods for further details. We can then compute the spin supercurrent from Eq. (\ref{eq:spinsuper}), and the charge supercurrent $I_Q$ can be obtained from the same formula by removing $\hat{\tau}$ from the trace and taking $I_{S_0}\rightarrow 2I_{S_0}e/\hbar=I_{Q_0}$, where $e$ is the electronic charge.

The critical charge supercurrent $I_Q^C$, obtained at a phase-difference $\phi=\pi/2$, is shown in Fig.~\textbf{\ref{fig:dFCur}a}, demonstrating that it becomes long-ranged even if there is no magnetic inhomogeneity and only a single ferromagnet is used. The physical mechanism behind this effect is that the spin-orbit coupling present in the thin, heavy normal metal layers rotates the triplet Cooper pairs due to an anisotropic spin relaxation \cite{BT14}. The existence of the long-ranged supercurrent in our setup requires a thin layer with spin-orbit coupling at each of the superconducting interfaces: with only one layer, the effect is absent. In this sense, the heavy normal metal layers effectively play an analogous role as the misaligned magnetization layers in the trilayered magnetic Josephson setup proposed in Ref. \cite{houzet_prb_07} as the spin-orbit coupling provides the spin-rotation required to produce the long-ranged triplet Cooper pairs. The key distinguishing aspect regarding the appearance of a long-ranged supercurrent in our work compared to previous proposals is that only one single homogeneous ferromagnet (without any intrinsic spin-orbit coupling) is required. Moreover, our analysis reveals new physical mechanisms at work in such structures, to be discussed below. 

The spin-orbit coupling is described by $\alpha$ and $\beta$, being respectively the Rashba and Dresselhaus coefficients. These are normalised to the superconducting gap $\Delta$ and length of normal metal $L_N$ in such a way that with a niobium superconductor of gap $\Delta\approx 3$~meV, $\alpha=0.5/L_N$ corresponds to a Rashba parameter of the order $3\times 10^{-12}$~eV m. It is clear from Fig.~\textbf{\ref{fig:dFCur}a} that the critical current decays rapidly in the absence of spin-orbit coupling ($\alpha=\beta=0$), and that this decay is strongly suppressed by the inclusion of spin-orbit coupling (note the log scale).

To model the ferromagnet, we assumed an exchange field $\bs{h}=h(0,\cos\theta,\sin\theta)$, with a strength $h/\Delta=50$ and a canting of $\theta=0.3\pi$ between the in- and out-of-plane components. The supercurrent exists for any orientation of the exchange field $\theta\in(0,\pi/2)$ and we will later discuss the precise dependence on the canting angle $\theta$. We choose $\tilde{G}_{M\! R}=0.2$ for the normalized interfacial magnetoresistance term and $\tilde{G}_\theta=1$ for the interfacial scattering phase shift on both sides \cite{eschrig_njp_15}. In this case, and with a typical superconducting coherence length of $\xi_S=25$ nm, the LR component dominates for ferromagnets of length $L_F$ greater than $\sim 10$ nm, causing the critical current $I_Q^C$ to decay slowly despite the presence of an exchange field $h\gg\Delta$, remaining orders of magnitude larger than the SR component for increasingly long ferromagnets. In this scheme, the associated current densities for a sample length $L_F\sim 10$ nm will be of the order $|j_Q^C|\sim 10^3$ A/cm$^2$ without spin-orbit coupling, and 1-2 orders higher with its inclusion (see Methods for details). This corresponds well with charge current densities measured in the experiment of Ref.~\cite{oboznov_prl_06}, which also used a CuNi alloy as the ferromagnet. For stronger exchange fields, the LR component will dominate for even shorter junctions, but the overall current magnitude will be suppressed. The supercurrent carried by the LR Cooper pairs can be significantly enhanced by including Dresselhaus coupling, as can be seen from the dotted line in Fig.~\textbf{\ref{fig:dFCur}a}, in which case the achievable critical charge current is much greater than with Rashba coupling alone.

We now turn to the spin supercurrent. Without spin-orbit coupling, no spin current flows in the junction. To demonstrate the physical origin of the dissipationless spin current and its different polarization components, it is useful decompose the triplet correlations in the system into their long-ranged and short-ranged contribution: $\bs{f} = \bs{f}_\text{LR} + \bs{f}_\text{SR}$. To take an explicit example, consider the case with pure Rashba coupling and an exchange field $\bs{h} = (0,h_y,h_z)$. In that case, we may write the general expressions: 
\begin{align}
\bs{f}_\text{LR} = (g_1, -g_2h_z/h, g_2h_y/h),\;\; \;  \bs{f}_\text{SR} = (0, f_\text{SR}h_y, f_\text{SR} h_z)/h,
\end{align}
so that $\bs{f}_\text{LR} \cdot \bs{h} = 0$ when $\bs{f}_\text{SR} \parallel \bs{h}$. Here, $\{g_1,g_2\}$ and $f_\text{SR}$ are complex scalars that describe the LR and SR parts of the superconducting correlations. Now, the spin expectation vector of a triplet Cooper pair is obtained by $\langle \vecS \rangle = \i \bs{f} \times \tf$ with $\tilde{f}(\varepsilon) = f^*(-\varepsilon)$. Inserting the long-ranged state $\bs{f}_\text{LR}$, one obtains 
\begin{align}
\langle \vecS \rangle_\text{LR} = -\i(\tilde{g}_2g_1 - g_2\tilde{g}_1)(h_y\hy + h_z\hz)/h.
\end{align}
This means that the spin of the LR Cooper pairs points along the exchange field, as expected. Similarly, one finds that $\langle \vecS \rangle_\text{SR} = 0$ for the SR Cooper pairs. However, there exists an additional contribution. The spin expectation vector of the total proximity-induced superconducting state may be written 
\begin{align}
\langle \vecS \rangle_\text{tot} &= \i (\vecf_\text{LR} + \vecf_\text{SR}) \times (\tf_\text{LR} + \tf_\text{SR}) \notag\\
&= \langle \vecS \rangle_\text{LR} + \i( \vecf_\text{LR} \times \tf_\text{SR}+ \tf_\text{LR} \times \vecf_\text{SR}).
\end{align}
It follows that there exists a novel \textit{interference term} $\langle \vecS \rangle_\text{int}=\i(\vecf_\text{LR} \times \tf_\text{SR} +\tf_\text{LR} \times \vecf_\text{SR})$ between the LR and SR Cooper pairs, which upon insertion of $\vecf_\text{LR}$ and $\vecf_\text{SR}$ produces two terms, $\langle \vecS \rangle_\text{int}=\langle \vecS \rangle_\text{ex}+\langle \vecS \rangle_\text{mix}$: 
\begin{align}
\langle \vecS \rangle_\text{mix} = S_1(\hy h_z - \hz h_y)/h,\;\; \;  \langle \vecS \rangle_\text{ex} = S_2\hx, \label{eq:ex}
\end{align}
where $S_j = -\i(\tilde{f}_\text{SR}g_j - f_\text{SR}\tilde{g}_j)$.

The \textit{exchange} term $\langle \vecS \rangle_\text{ex}$ of Eq.~(\ref{eq:ex}) is independent of the direction of the field $\bs{h}$. In contrast, $\langle \vecS \rangle_\text{mix}$ changes its spin-polarization direction as $\bs{h}$ is altered. We will explain the physical meaning of each of these terms in the section below. The critical spin supercurrent variation with $L_F$ is shown in Fig.~\ref{fig:dFCur}(b), displaying both the component parallel with the exchange field $(\bs{n} \parallel \bs{h}$), denoting it $I^C_{S,\parallel}$, and the magnitude of the perpendicular components
\begin{align}
|I^{C}_{S,\perp}| = \sqrt{(I_{S,\text{ex}}^{C})^2 +(I_{S,\text{mix}}^{C})^2}.
\end{align}
It is clear that the polarization of the spin supercurrent along the magnetization direction has a qualitatively different behavior with the length of the system compared with the polarization perpendicular to the exchange field, which oscillates within its typical exponential decay since it is limited by the penetration depth of the short-ranged superconducting correlations. Although both components decay exponentially, the penetration depth of the parallel component is enhanced greatly by the addition of spin-orbit coupling and it persists for significantly longer interstitial ferromagnets. Note that there is a non-monotonic relationship between the maximal supercurrents and the magnitude of the spin-orbit coupling, in the same way as there exists a non-monotonic relation between the density of states and spin-orbit coupling in a ferromagnet \cite{JacobsenLinder2015}. The local density of states would be expected to display a peak at zero energy whenever the long-range component of the spin supercurrent dominates in the system, and for there to be an increase in the critical temperature of the superconductor \cite{JOL}. Since the long-range correlations are carried by the so-called odd-frequency pairs \cite{AsanoTanakaGolubov2007,EschrigLofwander2008}, the system in this way reproduces features of unconventional superconductivity \cite{TanakaGolubov2007} using conventional $s$-wave superconductors.\\
\text{ }\\

\begin{figure*}
\includegraphics[width=\textwidth,angle=0,clip]{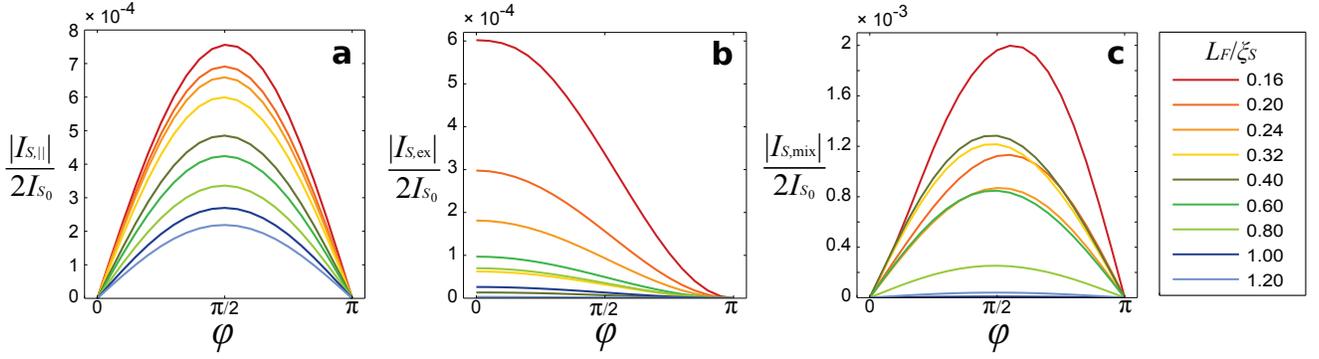}
\caption{\textbf{Controlling spin flow and polarization via superconducting phase difference}. The dependence of the spin supercurrent on the phase difference $\phi$ between the superconductors of the junction illustrated in Fig.~\ref{fig:model} is shown. The component parallel to the exchange field $\bs{h}=h(0,\cos\theta,\sin\theta)$ is given in \textbf{a}, the component perpendicular to the field polarized in the $x$-direction in \textbf{b} and the perpendicular component along $(0,\sin(\theta),-\cos(\theta))$ in \textbf{c}. The spin-orbit coupling is chosen to be of pure Rashba type with $\alpha=0.5/L_N$, and the parameters used are otherwise the same as in Fig.~\textbf{\ref{fig:dFCur}}. Results with both Rashba and Dresselhaus coupling are qualitatively similar, with consistently higher current magnitudes (not shown).}
\label{fig:phase}
\end{figure*}

\section*{Controlling spin polarization with the superconducting phase}

Analyzing the dependence of the spin supercurrent on the phase difference between the superconductors exposes another fundamental difference between the parallel and perpendicular components. We will prove that \textit{(i)} there exists a superconductivity-mediated exchange interaction in the system, even in the absence of any charge supercurrent and magnetic inhomogeneities, which acts with a torque on the magnetic order parameter, and that \textit{(ii)} both the magnitude and polarization direction of the spin supercurrent can be tuned via the superconducting phase difference. 

The phase-dependence of the component of the spin supercurrent parallel to the exchange field, $I_{S,\parallel}$,  is plotted in Fig.~\textbf{\ref{fig:phase}a}, and shows the expected first-order sinusoidal dependence on the phase difference $\phi$. This is physically reasonable since this component of the spin supercurrent is carried exclusively by the LR Cooper pairs which are polarized along the exchange field. When considering the perpendicular components of the spin supercurrent, however, the analysis in the preceding section showed that there exists two contributions $I_{S,\text{ex}}$ and $I_{S,\text{mix}}$ that originate from a novel interference between the LR and SR Cooper pairs. In order to unveil the physical meaning of these terms, we plot the variation of these with $\phi$ in Figs.~\textbf{\ref{fig:phase}b} and \textbf{c}. It is seen that these polarization components exhibit a fundamentally different response to the superconducting phase difference: $I_{S,\text{ex}}$ is invariant under time-reversal $\phi\to(-\phi)$ and finite even in the absence of any phase difference $\phi=0$ where no net charge current flows, whereas $I_{S,\text{mix}}$ is antisymmetric under time-reversal.
In effect, there exists a pure spin supercurrent flow without any charge current contamination in the system, even in the \textit{absence} of any magnetic inhomogeneities or half-metallicity.

Based on these observations, we offer the following interpretation of our findings. The polarization component of the spin supercurrent $\parallel$ $\bs{h}$ is understood simply as the polarization of the LR Cooper pairs that carry the long-ranged charge current and thus obeys the same type of current-phase relation as the charge current itself, vanishing both at $\phi=0$ and $\phi=\pi$. The interference between the SR and LR Cooper pairs now provides the spin supercurrent components with distinct physical origins. The term $I_{S,\text{mix}}$ represents the spin polarization that arises due to interference between LR and SR pairs carrying charge current, and is thus qualitatively similar to the charge current itself, with a $\sin\phi$ profile. In contrast, the term $I_{S,\text{ex}}$ represents something more exotic: \textit{it is a superconductivity-induced torque acting on the magnetization, which is present even in the absence of any charge current}. From its numerical evaluation, we find that it may be written as $|I_{S,\text{ex}}| = \mathcal{J}_1 + \mathcal{J}_2\cos\phi$, with the constants $\{\mathcal{J}_1,\mathcal{J}_2\}$ depending on system-specific details such as the strength of the exchange field $h$, the length of the ferromagnet $L_F$ and the strength of spin-orbit coupling $\alpha$. This means that the exchange spin supercurrent is invariant under $\phi \to (-\phi)$ and that it has a term that is independent of the superconducting phase difference.

\begin{figure*}
\includegraphics[width=\textwidth,angle=0,clip]{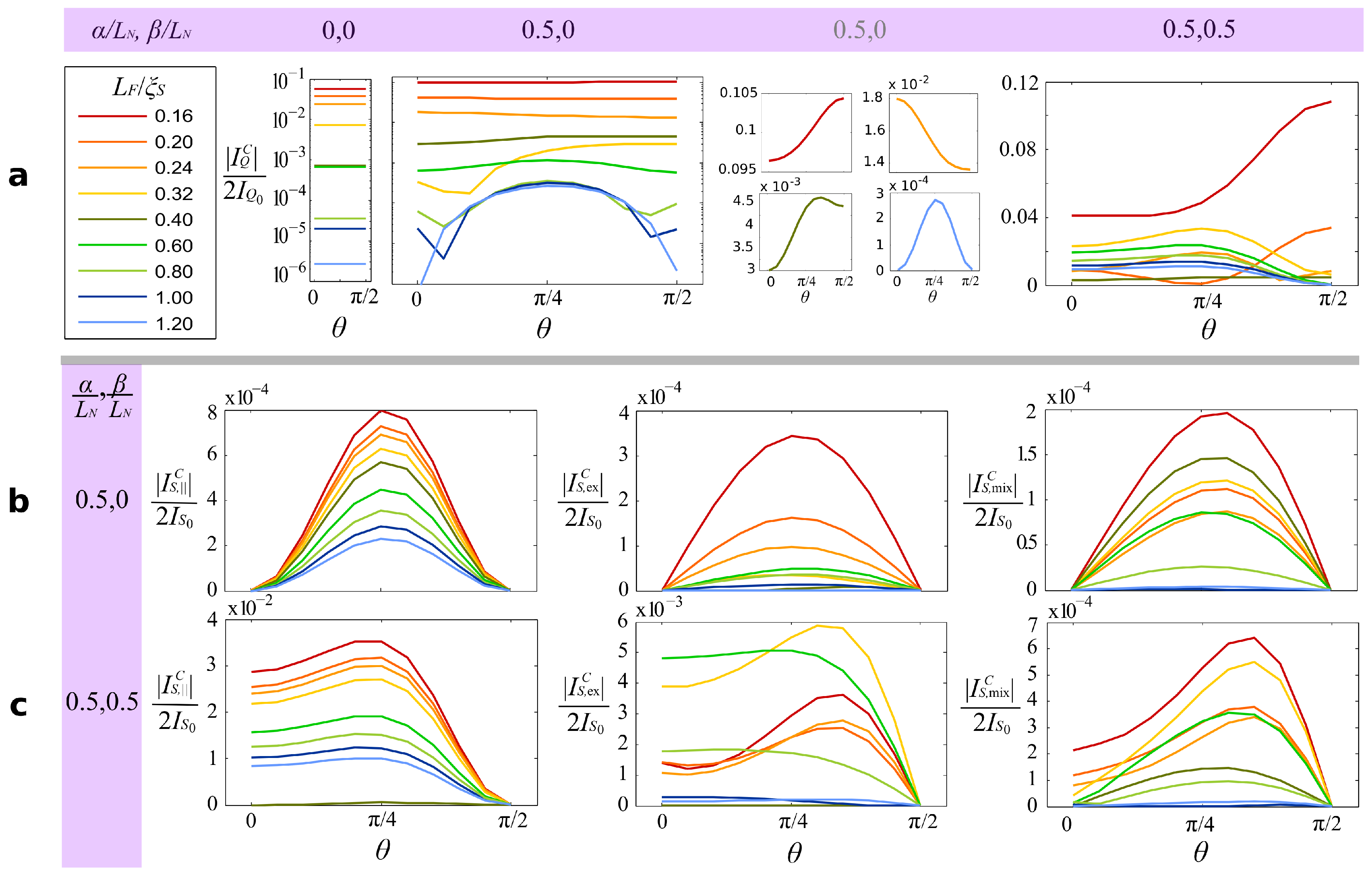}
\caption{\textbf{Charge- and spin-current vs. canting angle}. The effect of the canting angle $\theta$ between the in- and out-of-plane components of the exchange field $\textbf{\textit{h}}=50\Delta(0,\cos\theta,\sin\theta)$ is shown for the charge current in \textbf{a}, and for the spin-current components in \textbf{b} and \textbf{c}. Without spin-orbit coupling, the charge current does not depend on the magnetization orientation, and there is zero spin-current. With Rashba spin-orbit coupling we see a significant enhancement in the charge current, with a canting profile stabilising towards a sinusoidal maximum at $\theta=\pi/4$ for increasingly large ferromagnets as the long-ranged triplet component become dominant. The parallel component of the spin-current monotonically decreases with ferromagnet length, while the perpendicular components are sensitive to the $0$-$\pi$ transition in the ground state. The inclusion of Dresselhaus spin-orbit coupling yields a dramatic increase in both charge- and spin-current, and it is evident that purely in-plane magnetization $(\theta=0)$ is sufficient to generate the long-range component.}
\label{fig:canting}
\end{figure*}

The physical origin of this term is the following. Due to the proximity effect, both LR and SR superconducting correlations are induced in the ferromagnet in the presence of the inversion-symmetry breaking normal metal layers. The interference between these correlations create, according to Eq. (\ref{eq:ex}), a net spin moment. Since this moment is misaligned with $\bs{h}$, it acts with a torque on the magnetic order parameter $\bs{h}$, attempting to rotate it so that the net torque vanishes. The presence of magnetic anisotropy in the system could be expected to attempt to counteract this torque. Importantly, this effect is present even without any net charge flow $(\phi=0)$ and exists with just a single, homogeneous ferromagnet. This is evident by comparing Figs.~\textbf{\ref{fig:phase}b} and \textbf{c}, where the different polarization components of the spin supercurrent are plotted against the superconducting phase difference. This result shows that the magnitude and polarization direction of a dissipationless spin current can both be tuned exclusively via the superconducting phase difference, which is a surprising finding that offers a new way to control spin flow. The superconducting phase difference may itself be set in the conventional way via current-bias, or by applying an external magnetic flux in a loop-geometry\cite{leSueur2008}. We underline that this superconductivity-mediated exchange interaction is very different from exchange interactions in \eg conventional spin-valves with two ferromagnets, where a deviation from the parallel or antiparallel configuration produces a net equilibrium spin current that tries to align the magnetizations via a spin-torque \cite{slonczewski,slonczewski2, nogueira}. In contrast, here such a torque exists even with a single, homogeneous ferromagnet due to a unique interference effect between long-ranged and short-ranged triplet Cooper pairs.

It is clear from Fig.~\textbf{\ref{fig:phase}} that the maximal spin-current polarized along the exchange field is achieved around $\phi=\pi/2$, corresponding well with the definition of the critical spin current, taken to be the spin polarization of the critical charge current. These simulations were run for a canting angle of $\theta=0.3\pi$, and since this angle is in large part determined by material and geometry constraints it is instructive to consider the effect of the canting angle on the results. This is shown in Fig.~\textbf{\ref{fig:canting}}, and demonstrates that the long-ranged component of the charge current favours a canting angle of $\theta=\pi/4$, visible at longer sample lengths. It is also clear that the inclusion of both Rashba and Dresselhaus spin-orbit coupling allows the long-ranged component to be generated with a purely in-plane exchange field \cite{BT13,BT14}. \\

%%%%%%%%%%%%%%%%%%%%%%%%%%%%%%%%%%%%
%% 																		THIRD MAIN RESULT																 %%
%%%%%%%%%%%%%%%%%%%%%%%%%%%%%%%%%%%%

\section*{Spin-flip immunity}

Upon analysing the spin supercurrent in the above structure, one discovers an additional feature which pertains uniquely to currents generated by superconductors. Unlike conventional spin-polarized currents, we find that a spin \textit{super}current does not decay due to either spin-orbit impurity scattering or spin-flip scattering caused by magnetic impurities. This result has immediate implications for the usage of superconductors in spintronics, since it means that spin-flow created in this way is preserved even in regions with strong spin-flip scattering. We emphasize that this stands in complete contrast to conventional spin-currents, which have a decay length dictated by the amount of spin-flip scattering present.

Here we provide a general proof that the spin supercurrent is conserved both in normal metal and ferromagnetic systems, even in the presence of spin-orbit impurity scattering and isotropic spin-flip scattering from magnetic impurities. Using the relation between the Keldysh, retarded and advanced components of the Green function which holds at equilibrium (Eq.~(\ref{eq:equilib})), the Usadel equation may be written
\begin{align}\label{Eq:Usadel}
D \partial_z \text{Tr}\{\hat{\rho}_3 \hat{\tau}_j \hat{g}^R \partial_z \hat{g}^R \} + \i \text{Tr}\{\hat{\rho}_3 \hat{\tau}_j [\hat{\Sigma}, \hat{g}^R ]\}=0,
\end{align}
where we have defined 
\begin{align}
\hat{\Sigma} = \varepsilon\hat{\rho}_3 + \hat{M} - \hat{\sigma}_\text{so} - \hat{\sigma}_\text{sf},
\end{align} 
and $\hat{\tau}_j$ denotes the polarization-direction of interest. $\hat{M}=\textrm{diag}(\bs{h}\cdot\boldsymbol{\sigma},(\bs{h}\cdot\boldsymbol{\sigma})^*)$, where $\bs{h}$ is the magnetic exchange field, whereas the spin-orbit and magnetic impurity spin-flip self-energies have been included via the terms $\hat{\sigma}_\text{so}$ and $\hat{\sigma}_\text{sf}$ (see Methods for details). For any matrix $\hat{X}$ one has $\text{Tr}\{\hat{X}^\dag\} = (\text{Tr}\{\hat{X}\})^*$, from which it follows that if 
\begin{align}
\text{Tr}\{\hat{\rho}_3 \hat{\tau}_j [\hat{\Sigma}, \hat{g}^R ]\} = 0,
\end{align}
then the spin supercurrent will be conserved. By inserting the most general expression for the quasiclassical retarded Green function $\hat{g}^R$ [given in Eq.~(\ref{Eqn:gR})], direct evaluation shows that the above trace is always zero in the absence of an exchange field despite the presence of spin-flip scattering. In the presence of an exchange field, the same holds for the spin supercurrent $I_{S,\parallel}$ polarized along the magnetization and remains true even if the exchange field is spatially inhomogeneous. Even though the magnitude of the spin supercurrent is reduced with increasing spin-flip scattering \cite{gomperud_prb_15}, it is remarkable that a spin supercurrent, controllable via the superconducting phase difference, has no decay even if both spin-orbit and magnetic impurities are present in the sample. \\

\section*{Summary}

In conclusion, we have shown three major results: (i) a long-ranged spin supercurrent can be created without any magnetic inhomogeneities, (ii) both the magnitude and polarization direction of the spin supercurrent can be tuned separately via the superconducting phase difference, and (iii) spin supercurrents created in this way do not decay even in the presence of spin-flip scattering, \ie they display spin-flip immunity. We have proposed that this may be observed experimentally in a Josephson junction consisting of conventional $s$-wave superconductors (\eg Nb) with very thin layers of a heavy normal metal (\eg Pt or Au) and a single homogeneous ferromagnet with magnetocrystalline out-of-plane anisotropy (\eg PdNi or CuNi). We would like to note that no ``exotic" materials, such as unconventional superconductors or noncentrosymmetric ferromagnets, are required -- the effects predicted in this work appear by combining conventional superconductors and metals, which should make experimental verification of our results readily achievable. Our results confirm the significant and immediate advantage that superconductors offer spintronics. \\

\section*{Methods}

We solved the Riccati parameterised Usadel equation with spin-orbit coupling \cite{JOL} iteratively between the layers, using the NOTUR supercomputer facilities (Kongull), for the full proximity effect regime. Having three separate layers between the superconductors means the setup is not amenable to analytic solution even in the weak proximity limit, in contrast to several other works cited in the main text. In the normal metal, spin-orbit coupling is included in the Usadel equation Eq.~(\ref{Eq:Usadel}) by replacing the derivative with its covariant equivalent. We describe the normal-metal-ferromagnet interfaces via the spin-dependent boundary conditions
\begin{align}
2L_j\zeta_j\hat{g}_j\partial_z\hat{g}_j=&\left[\hat{g}_j,\hat{g}_k\right]+2L_j\zeta_j\hat{g}_j \i \left[\hat{\bs{A}}_z,\hat{g}_j\right]\nonumber\\
&+\sigma_j \tilde{G}_{M\! R}\left[\hat{g}_j,\left\{\hat{M},\hat{g}_k\right\}\right]+\sigma_j \i\tilde{G}_\theta\left[\hat{g}_j,\hat{M}\right],
\end{align}
where $j,k=\{\textrm{left},\textrm{right}\}$, $j\neq k$ denotes the two sides of the interface and the orientation determines the sign $\sigma_{\textrm{right}}= 1$, $\sigma_{\textrm{left}}= -1$. The thin-film layering direction is taken to be in the $z$-direction, and $\tilde{G}_{M\! R}$ and $\tilde{G}_{\theta}$ denote the interfacial magnetoresistance and scattering phase shifts respectively. We chose $\zeta_j=3$ for the transparency parameter of all interfaces. The spin-orbit coupling field $\hat{\bs{A}}=\textrm{diag}(\bs{A},-\bs{A}^*)$, and we have considered the case $\bs{A}=(\beta\sigma_x-\alpha\sigma_y,\alpha\sigma_x-\beta\sigma_y,0)$, where $\alpha, \beta$ are the Rashba and Dresselhaus coefficients respectively. Note that if the spin-orbit field contains a component along the junction direction, for example if isolating the triplet component via a $\pi$-biased junction \cite{JacobsenLinder2015}, then the relative sign of the spin-orbit coupling between the two normal layers becomes important. The extrinsic spin-orbit scattering and spin-flip terms are given by 
\begin{align}
\hat{\sigma}_{so}&=-\frac{1}{8\tau_{so}}\sum_i\hat{\alpha}_i\hat{\rho}_3\hat{g}^R\hat{\rho}_3\hat{\alpha}_i,\nonumber\\
\hat{\sigma}_{s\! f}&=-\frac{1}{8\tau_{sf}}\sum_i\hat{\alpha}_i\hat{g}^R\hat{\alpha}_iS_i,
\end{align}
where $\tau_{so}$ and $\tau_{sf}$ are the mean scattering times, $S_i$ is the spin expectation value and we have defined the matrix $\hat{\alpha}_i=\textrm{diag}(\sigma_i,\sigma_i^T$). The general form of the retarded Green function is
\begin{eqnarray}\label{Eqn:gR}
\hat{g}^R=
\begin{pmatrix}
N(\textit{I}+\gamma\tilde{\gamma}) & 2N\gamma\\
-2\tilde{N}\tilde{\gamma} & -\tilde{N}(\textit{I}+\tilde{\gamma}\gamma)
\end{pmatrix},
\end{eqnarray}
with normalization matrices $N=(\textit{I}-\gamma\tilde{\gamma})^{-1}$ and $\tilde{N}=(\textit{I}-\tilde{\gamma}\gamma)^{-1}$ and identity matrix $I$. The $\tilde{\cdot}$ operation denotes complex conjugation and $\varepsilon\rightarrow(-\varepsilon)$. Regarding the choice of junction parameter, one may consider a reasonable approximation of the normal-state density of states to be of the order $N_0\sim 10^{22}/$(eV cm$^3$), and the diffusion constant of CuNi to be\cite{oboznov_prl_06} $D\sim 5$ cm$^2$/s. Moreover, in Figs.~\textbf{\ref{fig:dFCur}} and \textbf{\ref{fig:canting}} we have shown the critical charge current and spin currents,
computed at a phase difference of $\phi=\pi/2$. The critical charge
current may deviate slightly from this phase difference near the
transition points between the 0 and $\pi$ ground states since higher
order harmonics may become increasingly significant when the
current is very small, but this is largely negligible for our scheme.

\text{ }\\

\section*{Acknowledgements}

We thank J.A. Ouassou for useful discussions on computational aspects as well as N. Banerjee, J. Moodera, A. Di Bernardo, and J. W. A. Robinson for helpful comments. We acknowledge funding via the ``Outstanding Academic Fellows'' programme at NTNU, the COST Action MP-1201 and the Research Council of Norway Grant numbers 205591, 216700, and 240806.\\

\section*{Author contributions statement}

S.H.J. performed the full proximity analysis and prepared the figures. I.K. carried out initial investigations in the weak proximity limit.  J.L. had the idea and supervised the project. All authors analyzed the results and contributed to writing the manuscript.\\

\section*{Additional information}

\textbf{Competing financial interests} The authors declare no competing financial interests.

\end{document}